\documentclass[aps,prb,twocolumn,footinbib]{revtex4}         
\usepackage{amsmath}
\usepackage{amssymb}
\usepackage{mathrsfs}
\usepackage{graphicx}

\newcommand{\ab}{{\em ab initio}\ }

\newcommand{\mone}{$\bar{1}$}
\newcommand{\dg}{$^\circ\mathrm{C}$}
\newcommand{\beg}{\begin{equation}}
\newcommand{\e}{\end{equation}}
\newcommand{\G}{\mathit{\Gamma}}
\newcommand{\bi}{B$_i$}
\newcommand{\bs}{B$_s$}
\newcommand{\bd}{B$_d$}

\newcommand{\Gseg}{G_{\mathrm{seg}} }
\newcommand{\Eseg}{E_{\mathrm{seg}} }

\newcommand{\Wsep}{W_{\mathrm{sep}} }
\newcommand{\A}{\textbf{A}}
\newcommand{\B}{\textbf{B}}
\newcommand{\C}{\textbf{C}}
\newcommand{\D}{\textbf{D}}
\newcommand{\Am}{\mathbf{A}}
\newcommand{\Bm}{\mathbf{B}}

\newcommand{\As}{\textbf{A}$_\mathbf{s}$}
\newcommand{\Bs}{\textbf{B}$_\mathbf{s}$}
\newcommand{\Cs}{\textbf{C}$_\mathbf{s}$}
\newcommand{\Ds}{\textbf{D}$_\mathbf{s}$}

\newcommand{\Bsm}{\mathbf{B}_\mathbf{s}}

\newcommand{\Gcl}{$\mathcal{G}_\mathrm{cleav}$}
\newcommand{\Gdl}{$\mathcal{G}_\mathrm{disl}$}
\newcommand{\Gclm}{\mathcal{G}_\mathrm{cleav}}
\newcommand{\Gdlm}{\mathcal{G}_\mathrm{disl}}
\def\Jm{J/m$^{2}$}

\begin{document}

\title{Boron in copper: a perfect misfit in the bulk and cohesion enhancer at a grain boundary} 

\author{A. Y. Lozovoi and  A. T. Paxton}
\affiliation{Atomistic Simulation Centre, School of Mathematics and Physics,
Queen's University  Belfast, Belfast BT7 1NN, Northern Ireland, U.K.}

\date{\today}

\begin{abstract}
Our \ab\ study suggests that boron segregation to   
the $\Sigma 5$(310)[001] grain boundary should strengthen the boundary up to 
1.5~ML coverage (15.24 at/nm$^2$). The maximal effect is observed at 0.5~ML and corresponds to boron
atoms filling exclusively grain boundary interstices.  
In copper bulk, B causes significant distortion both in interstitial and regular
lattice sites for which boron atoms are either too big or too small.    
The distortion is compensated to large extent when the interstitial and substitutional boron
combine together to form a strongly bound dumbell. Our prediction is that 
bound boron impurities should appear in sizable proportion
if not dominate in most experimental conditions. A large discrepancy between calculated
heats of solution and experimental terminal solubility of B in Cu is found, indicating
either a sound failure of the local density approximation or, more likely, strongly 
overestimated solubility limits in the existing B--Cu phase diagram.  
\end{abstract}

\maketitle

\section*{Introduction}

Boron has an extremely good record in improving intergranular cohesion in metals.
It is mostly famous for curing the long standing problem of room temperature brittleness 
in Ni$_3$Al.\cite{liu85}  
Boron segregation was found to reinforce grain boundaries in other intermetallic 
compounds (FeAl, NiAl, Ni$_3$Si) and to improve low temperature ductility in  
bcc iron and refractory metals, such as Mo and W (see, {\it e.g.}, 
Ref.~[\onlinecite{lejcek03}] and references therein). 

The effect of boron addition on copper is far less studied. According to the Cu--B phase  
diagram,\cite{Massalski} boron solubility in copper is low, 0.06 at.\% at 
room temperature rising to 0.29 at.\% at the eutectic temperature 1013~\dg. 
Dissolved boron has a strong propensity to segregate to surfaces and interfaces. 
It is not clear whether segregation weakens or strengthen grain boundaries. 
Nevertheless, doping copper with boron is found to be efficient in preventing 
segregation of antimony to grain boundaries.\cite{glikman74}
Substantial improvement of mechanical properties of \textit{nanocrystalline} Cu 
samples is reported as B segregation can be used to limit grain growth
during heat treatment.\cite{sur99} Despite such encouraging experimental findings, 
quite surprisingly, no theoretical simulations of boron at copper 
grain boundaries or even free surfaces, are known to us. 

Similarly poor is the situation with studies of boron in bulk copper.
It is not even clear whether boron occupies interstitial or substitutional positions.
Analysis of boron's neighbours in the Periodic Table does not rule out either
possibility. Carbon is an interstitial impurity in Cu,\cite{ellis99} whereas Al and Be are 
substitutional impurities. In recent experimental work in which accelerated boron ions were 
implanted in Cu, both types of impurities were observed.\cite{fullgrabe01} 

In the present study we employ standard density functional calculations to study the 
behaviour of boron impurities  at a copper grain boundary and in the bulk. 
We find that boron \textit{strengthens} the   
$\Sigma 5$\{310\}[001] symmetric tilt grain boundary in the whole range of boundary coverages 
investigated (up to 1.5~ML). The maximum strengthening occurs at 0.5~ML at which boron exclusively 
occupies grain boundary interstices. We further identify mechanisms responsible for  
grain boundary strengthening within the framework of  the ``ghost impurity cycle'' proposed in 
our previous work.\cite{lozovoi06} The cycle admits the occupation of both substitutional and 
interstitial positions by impurity atoms at the interface, a feature that is fully exploited
in the present study.   

Due to the peculiar interplay of atomic sizes, interstitial and substitutional positions are equally
unwelcoming to boron in bulk Cu. Boron is too big an interstitial impurity and too small 
a substitutional impurity. As a result, both have nearly the same heat of solution with the 
interstitial position marginally preferred. If, however, B atoms combine in dumbells then most of the elastic 
distortion of the host is eliminated and significant lowering of the heat of solution is achieved.   
The energy gain is so large that boron dumbells should persist up to high temperatures.
Another surprising finding of our study is a large disagreement between the theoretical heat of 
solution and experimentally observed maximal solubility of B in Cu.   
 
The paper is organised as follows. Sec.~\ref{sec_cycle} outlines definitions and thermodynamical 
relations used in the present work.  The computational setup is described in Sec.~\ref{sec_calc}.   
The results on B in bulk Cu are presented in Sec.~\ref{sec_bulk}, 
whereas Sec.~\ref{sec_gb} is concerned with the effect of boron at 
the grain boundary. In the latter section, we first look at the change of the 
work of separation and at the impurity 
segregation energies (Sec.~\ref{subsec_wsep}), then we discuss 
atomic structure of the boundaries and free surfaces with different boron 
content (Sec.~\ref{subsec_astruc}), and finally apply the ``ghost impurity 
cycle'' to reveal the mechanisms responsible for the cohesion enhancement 
(Sec.~\ref{subsec_bmech}). Our main findings are summarised in Sec.~\ref{sec_concl}. 
A thermodynamic model used to estimate the concentration of different kinds of 
boron impurity in bulk copper, is outlined in the Appendix.

\section{Work of separation and ghost impurity cycle}
\label{sec_cycle}

\begin{figure}
\begin{center}
\includegraphics[width=8.0 cm,angle=0]{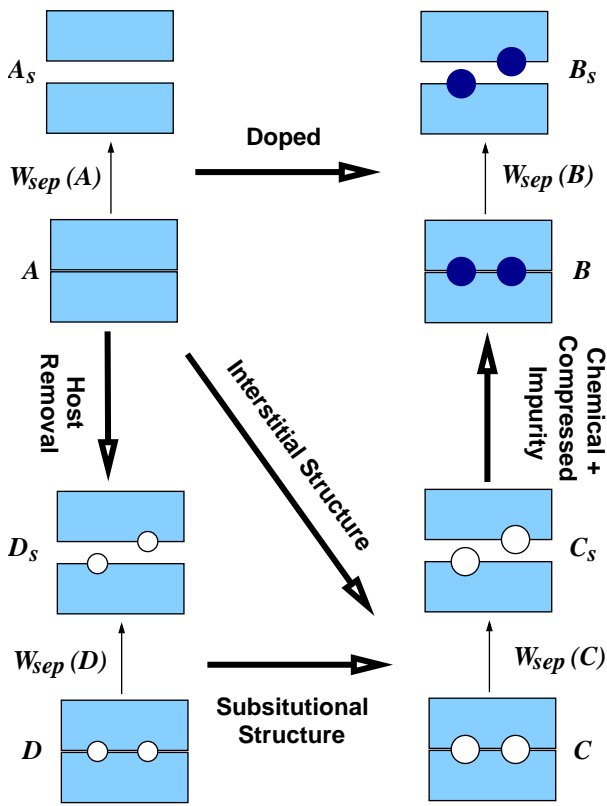}
\end{center}
\caption{``Ghost impurity cycle'' used for discussion of 
the effect of impurity on grain boundary strength.
\A\ and \B\ denote the pure and
segregated grain boundaries, respectively, in their equilibrium
geometry.
Boundary \C\ is created by substituting impurity atoms
in \B\ (black spheres) with vacancies (white spheres)
without further atomic relaxation. Configuration \D\ is \A\ in 
which the host atoms that will be replaced by impurity in \B\,
are removed with other atoms kept in place. \As\ and \Bs\ denote the
free surfaces into which  grain boundaries \A\ and \B\ cleave.
\Cs\ and \Ds\ are prepared from \Bs\ and \Cs\, respectively, using the 
above strategy, namely, impurity atoms in \Cs, or host atoms to be
replaced by impurity in \Ds, are removed keeping the positions of other
atoms fixed. 
Path \A$\to$\D$\to$\C$\to$\B\ refers to substitutional impurities,
whereas path \A$\to$\C$\to$\B\ applies to interstitial impurities 
(Ref.~\onlinecite{lozovoi06}).
}
\label{fig_paths}
\end{figure}

The energy release rate (the minimal energy per unit area of crack advance) associated with 
brittle cleavage of a grain boundary, \Gcl, is the central quantity characterising the resistance  
of the boundary to decohesion in the Rice--Thomson--Wang approach.\cite{rice1974,RiceWang} 
If \Gcl\ is lower than the energy release rate associated with emitting one dislocation, \Gdl, 
then cracks remain atomically sharp and the crystal breaks in brittle manner. If $\Gclm > \Gdlm$,
the crack blunts and the crystal is ductile. Impurity segregation to grain boundaries 
can either decrease or increase \Gcl. Bi in Cu is a classic example of the former. Copper
grain boundaries with bismuth segregating eventually reach the condition $\Gclm < \Gdlm$ leading to a
ductile-to-brittle transition.   

In the limit of \textit{fast separation} which we assume in the present study, any impurity 
exchange between newly formed surfaces and bulk during the decohesion is prevented.
In this limit, \Gcl\ can be identified with the reversible \textit{work of separation} 
\beg
\label{Wsep}
\Gclm = \Wsep = \frac{1}{A} \left\{G^s - G^{gb} \right\},
\e
where $A$ is the surface area, $G^{gb}$ is the excess Gibbs free energy of
a representative piece of material containing grain boundary, and
$G^s$ is the sum of two excess Gibbs free energies corresponding to surfaces
formed after decohesion. Eq.~(\ref{Wsep}) assumes that the impurity excess $\G$ at 
the grain boundary is equal to the sum of surface impurity excesses. 
The excess in Eq.~(\ref{Wsep}) is defined with respect to the underlying bulk 
crystal. 

Our \ab\ calculations refer to the zero temperature limit hence the Gibbs free energies
are replaced with total energies. It is convenient then to express the changes in $\Wsep$
due to impurity
in terms of segregation energies per impurity atom, $\Eseg$. (Segregation energy 
is the energy required to remove all impurity from an interface and distribute it
in the bulk). Eq.~(\ref{Wsep}) becomes (\textit{cf.} Eq.~(6) in Ref.~\onlinecite{lozovoi06}):   
\beg
\label{Wseg}
\Wsep(\Bm) = \Wsep(\Am) + \G \left\{\Eseg(\Bm) - \Eseg(\Bsm)\right\} \, ,
\e
where \A\ and \B\ denote pure material and material with segregant, respectively.
Segregation energies are easy to obtain in an \ab\ supercell approach using the total
energies of supercells containing
grain boundary (or surface) with and without impurity, $E_{tot}(\Bm)$ and $E_{tot}(\Am)$, 
and the same combination for the bulk, $E_{tot}^{b}(\Bm)$ and $E_{tot}^{b}(\Am)$. 
For a grain boundary, for example, we have:
\begin{eqnarray}
\label{Eseg}
\Eseg(\Bm) &=& \frac{1}{N^{gb}} \left\{ E_{tot}^{gb}(\Bm) - E_{tot}^{gb}(\Am) \right\} 
\nonumber \\ &&
- \frac{1}{N^{b}} \left\{E_{tot}^{b}(\Bm) - E_{tot}^{b}(\Am)\right\} \, ,
\end{eqnarray}
where $N^{gb}$ and $N^{b}$ denote the number of impurity atoms included in grain boundary and 
bulk supercells, respectively. 
Eq.~(\ref{Eseg}) conveys a simple picture in which an impurity atom at the boundary is exchanged 
with a host atom in the bulk; if interstitial positions are involved then Eq.~(\ref{Eseg})
should by augmented by adding or subtracting a suitable amount of chemical potentials of the host
atoms which, again, are the total energies per atom in pure bulk.

\begin{figure*}[t]
\begin{center}
\includegraphics[width=11 cm,angle=0]{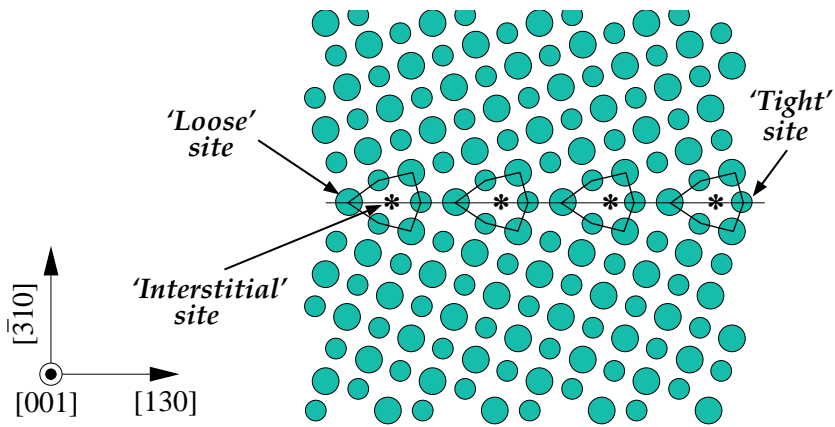}
\end{center}
\caption{Equilibrium structure of the $\Sigma 5$\{310\}[001] symmetric tilt grain boundary
in pure copper.  Larger and smaller circles represent alternating (001) Cu planes. 
The grain boundary plane contains two inequivalent Cu atoms to which we 
refer as to the ``loose'' and ``tight'' sites. The ``interstitial'' site suitable for 
segregation of small impurity atoms shown with an asterisk. 
}
\label{fig_pure}
\end{figure*}

Implicit in Eq.~(\ref{Eseg}) is that the bulk is sufficiently dilute so that neither interface
nor bulk supercells include any additional randomly distributed impurity atoms. In the dilute
limit, the grain boundary impurity excess $\G$ and excess volume $v^{xs}$ per unit area can
be found as  
\begin{eqnarray}
\G &=& \frac{N^{gb}}{A}  \label{Gamma} \\ 
v^{xs} &=& \frac{1}{A} \left\{V(\Bm) - N^{gb}_0 \Omega_0 \right\} \label{vxs}  \, ,
\end{eqnarray}
where $V(\Bm)$ is the volume of a relaxed grain boundary supercell with impurity, $N^{gb}_0$ is the number 
of host atoms in this supercell and $\Omega_0$ is the atomic
volume in pure bulk.

To separate various aspects of grain boundary  weakening or strengthening by impurity atoms originating
from their size, positions, and chemical identity, the ``ghost impurity cycle'' introduced     
in Ref.~\onlinecite{lozovoi06} and shown in Fig.~\ref{fig_paths}, is rather useful. In this cycle, 
the direct transition from unsegregated
to segregated state \A$\to$\B\ is replaced with a \textit{gedanken} path through intermediate
configurations \A$\to$\D$\to$\C$\to$\B\ for both grain boundary and surfaces, and the 
respective changes in $\Wsep$ are evaluated. Configuration \C\ is created from \B\ by removing
all impurity atoms without subsequent relaxation. These missing atoms are referred to as ``ghosts''   
since they create forces which keep host atoms in place but do not contribute to the energy of the
system in any other way. Such ``ghosts'' are distorted vacant sites for substitutional impurities or
centres of expansion for interstitial impurities.  Configuration \D\ is constructed from \A\ in
a similar way except that we remove the host atoms defined by impurity sites in \B\ only if these  
impurities replace host atoms. For interstitial impurities, configuration \D\ is not visited 
(see interstitial path in Fig.~\ref{fig_paths}). 

The same approach is used for the generation of surface configurations \As--\Ds.  
\Bs\ and \As\ represent the equilibrium geometry of free surfaces with and without impurity, respectively.      
Configuration \Cs\ is created from \Bs\ by removing impurity atoms,    
whereas in \Ds\ one removes only those host atoms that will be replaced with impurity in \Bs.
The remaining atoms in \Cs\ and \Ds\ are not allowed to move in response to removal of some of their 
neighbours. Note that the impurity atoms occupying substitutional 
positions at a grain boundary can be interstitial impurities at free surfaces (or \textit{vice versa}).  
In terms of Fig.~\ref{fig_paths}, this would mean that the substitutional path should be used for grain 
boundaries, whereas the interstitial path should be taken for surfaces. More generally, one
could envisage a situation in which \text{part} of impurity atoms occupy interstices whereas    
the other part substitute host atoms. We shall describe shortly how to deal with such situations.

As argued in Ref.~\onlinecite{lozovoi06}, the change of $\Wsep$ at the \A$\to$\D\ step
describes grain boundary weakening due to some host-host bonds being broken (``host removal''
mechanism, HR). Transition \D$\to$\C\ corresponds to the distortion of the atomic structure 
of pure boundary and surface caused by impurity (``substitutional structure'' mechanism, SS). 
Finally, step \C$\to$\B\ brings in the impurity-host chemical interactions and, if relevant, 
associated changes in neighbouring host-host bonds. For oversized impurity atoms, this step
also incorporates the elastic energy stored in compressed impurity atoms. These two mechanisms
can not be separated hence we refer to this step as ``chemical + compressed impurity'' mechanism (CC).          
For more details regarding the cycle and its implementation, the reader is referred to 
Ref.~\onlinecite{lozovoi06}.

The ``ghost impurity cycle'' outlined above treats 
interstitial and substitutional impurities on an equal basis. This is vital for 
the purposes of the present study in which we shall be introducing boron into substitutional
and interstitial positions at a grain boundary, sometimes even simultaneously.   
The way to deal with the ``co-existence'' of the substitutional and
interstitial paths in Fig.~\ref{fig_paths} is to formally include configuration \D\ into
the latter making it indistinguishable from \A. In such case Eq.~(\ref{Wseg})
can be used without making any specific allowances. The same applies to surfaces with 
impurities occupying adatom positions. These can be treated similarly to interstitial 
impurities at grain boundaries.

\begin{table*}[t]
\caption{Ground-state properties of $\alpha-$B: equilibrium volume, $V_0$, rhombohedral lattice constant, $a_0$,
rhombohedral angle, $\phi$, internal coordinates of boron atoms, $x_1, z_1, x_2,$ and $z_2$, bulk modulus, $B$,
and the energy difference (per atom) between fcc and $\alpha$ boron, $\Delta E_{fcc}$, 
obtained in the present study and other \ab\ calculations.
Experimental data in the last column are from Ref.~\onlinecite{donohue} unless stated otherwise.} 
\begin{tabular}{llccccccccccc}
\hline
\hline
Method                     &FP LMTO&PAW&PW-PP&LMTO&FP LMTO&PW-PP&PW-PP&PW-PP&PW-PP&US-PP&US-PP
& Exp. \\
LDA or GGA?                &\multicolumn{2}{c}{LDA}&\multicolumn{2}{c}{LDA}&\multicolumn{2}{c}{LDA}
                                                                    &LDA    &LDA & LDA &  GGA & GGA \\
Reference                  &\multicolumn{2}{c}{present study}
&\multicolumn{2}{c}{Ref.~\onlinecite{mailhiot90}}
&\multicolumn{2}{c}{Ref.~\onlinecite{barbee97}}
& Ref.~\onlinecite{lee90}
& Ref.~\onlinecite{vast97}
& Ref.~\onlinecite{masago06}
& Ref.~\onlinecite{prasad05}
& Ref.~\onlinecite{haussermann03}
& Ref.~\onlinecite{donohue}\\
\hline
$V_0$ (\AA$^3$)       & 6.899 &  6.993  & 7.05 & 6.93 & 7.06 & 6.88 &       &    &     & 6.946&7.30
& 7.337 [\onlinecite{nelmes93}]\\
$a_0$ (\AA)           & 4.967 &  4.989  &      &      &      &      & 5.034 &4.98&4.967& 4.973&    
&                          \\
$\phi$ (deg.)         & 58.055&  58.063 &      &      &      &      & 58.119&58.2&58.65&      &    
&58.06  \\
$x_1$                 & 0.0106& 0.0105  &     &      &0.010 &      &       &    &     &      &    
&0.0104 \\
$z_1$                 &--0.3460& --0.3458&     &      &--0.344&     &       &    &     &      &    
&--0.3427\\
$x_2$                 & 0.2211 & 0.2215 &    &      &0.220 &      &       &    &     &      &    
& 0.2206 \\
$z_2$                 &--0.3694& --0.3700&    &      &--0.369&     &       &    &     &      &    
&--0.3677\\
$B$ (GPa)             & 232   &         & 249  & 266  & 230  & 227  &       &    &218.4&      &    
&224 [\onlinecite{nelmes93}] \\
$\Delta E_{fcc}$ (eV) & 1.30  & 1.35    & 1.43 & 1.83 & 1.31 & 1.39 &       &    &     &      &
&\\
\hline
\hline
\end{tabular}
\label{tbl_balpha}
\end{table*}

\section{Calculation details}
\label{sec_calc}

The $\Sigma 5$\{310\}[001] tilt grain boundary is represented in our study 
by a periodic supercell containing two grain boundaries with opposite orientation, 
without any vacuum. Altogether, there are 38 atoms in the supercell, with each atom
representing one $\{310\}$ layer, except the grain boundary plain which contains two 
atoms, the ``tight'' and the ``loose'' sites. These two sites together with an
``interstitial'' site shown in Fig~\ref{fig_pure}, are considered as three possible 
segregation sites for B. Occupation of any one, any two, or all three of these sites 
corresponds to 0.5, 1, and 1.5~ML coverage in our notation. As we shall see in Sec.~\ref{subsec_astruc},
severe atomic relaxation of the boundary involving lateral translation of the grains
may significantly change the local environment of segregated atoms. We therefore shall
be labelling configurations with respect to positions in which the impurity was \textit{initially}
placed.      
  
To represent free (310) surfaces we use the same supercell with 25 layers of copper, 
the rest being vacuum. Outermost copper layers are replaced with impurity
layers if we need to model a segregated surface.  The situation in which a grain 
boundary containing 0.5 or 1.5 ML of impurity cleaves into surfaces with even amount of impurity 
requires us to double the supercell along the [001] direction. 
Cubic supercells containing up to 108 atoms (3$\times$3$\times$3 fcc cells) were used to model 
boron impurities in bulk Cu.

Our first principles calculations employed the full potential 
LMTO method as implemented in the NFP code.\cite{nfp} Calculations were semirelativistic, without 
spin polarisation. We used the local density approximation (LDA) 
in the parameterisation of von Barth and Hedin, modified by Moruzzi {\it et al.}\cite{vBH,MJW}
Other parameters ($k$-point meshes, real space meshes, etc.) were the same as in Ref.~\onlinecite{lozovoi06}
to which the reader is referred for further details. 
\section{Boron impurity in bulk copper}
\label{sec_bulk}

\subsection{Pure boron}

Solid boron can exist in a number of relatively stable allotropic 
modifications---rhombohedral, tetragonal, and even amorphous. 
It is not clear which phase corresponds to the ground state of B at ambient conditions. 
Two rhombohedral phases, $\alpha-$ and $\beta-$B are the most likely candidates.  
$\alpha-$B becomes unstable at 1200\dg\ and converts to $\beta-$B above 1500\dg, 
but $\beta-$B does not transform back to $\alpha-$B upon cooling (see [\onlinecite{lee90}] 
and references therein). Hence kinetic effects must impose severe limitations in this material. 
$\alpha-$B was found lower in energy by 0.036 eV/atom in Ref.~\onlinecite{masago06} and by
0.283 eV/atom in Ref.~\onlinecite{prasad05}, and
is assumed to represent the ground state in the present study.

The unit cell of the rhombohedral $\alpha-$B consists of twelve atoms forming an 
icosahedron. Equilibrium structure, bulk modulus and 
the fcc$-\alpha$ energy difference obtained in our study are in good agreement 
with other calculations and show the usual discrepancies with experiment associated with 
the LDA, namely, underestimation of atomic volumes and consequent overestimation of the
bulk moduli  
(see Table~\ref{tbl_balpha}). Calculations are scalar-relativistic,  
fully relaxed, and employ the 8$\times$8$\times$8 Monkhorst-Pack mesh of $k$ points.
Increasing the $k$ point mesh to 12$\times$12$\times$12 changes  
the total energy by less than 10$^{-5}$ Ry, whereas the forces remain within the 
convergence criterion 10$^{-3}$ Ry/Bohr used throughout the whole study. 
 
\begin{table}
\caption{Boron in bulk Cu: enthalpy of solution $H_s$ (per impurity atom) and the relative dilation 
volume $\Omega_d/\Omega_0$
of interstitial boron B$_i$, substitutional boron B$_s$, and boron dumbells B$_d$ with different orientation. 
$\Delta H_s$ is the enthalpy relative to that of the $s\langle100 \rangle$ dumbell (per entity).    
$N_{at}$ denotes the number of lattice sites in a supercell used in the calculation, 
$\Omega_d$ is the change of the volume of this supercell when either a single impurity 
(B$_i$ or B$_s$) or a dumbell B$_d$ is introduced, $\Omega_0$ is the atomic volume 
in pure fcc Cu, and $E_a$ is the activation energy for impurity diffusion. 
Experimental data on $E_a$ are from Ref.~\onlinecite{fullgrabe01}.
Theoretical 
results are obtained by the FP LMTO method unless indicated otherwise.  
}
\begin{tabular}{lccccc}
\hline
\hline
Impurity & $N_{at}$ & $H_s$, eV &  $\Delta H_s$, eV& $\Omega_d/\Omega_0$ & $E_a$, eV \\
\hline
B$_i$                         &    32 & 1.58 & $-$0.10     &     0.88  &  1.31 \\
B$_i^\dagger$                 &    32 & 1.63 & \ldots   &     0.93  & \\
B$_i$                         &   108 & 1.66 & 0.06     &     0.89  &  0.93 \\
B$_s$                         &    32 & 1.70 & 0.02     &  $-$0.44  & \\
B$_s$                         &   108 & 1.70 & 0.10     &  $-$0.45  & \\
B$_{d}$:\\ 
$a\langle111 \rangle$         &    32 & 1.69 & 1.70     &     0.52  & \\
$s\langle111 \rangle$         &    32 & 1.04 & 0.40     &     0.38  & \\
$a\langle100 \rangle^\ddagger$&    32 &\ldots&\ldots    &    \ldots & \\
$s\langle100 \rangle$         &    32 & 0.84 &   0      &     0.25  & \\
$s\langle100 \rangle$         &   108 & 0.80 &   0      &     0.24  & \\
Exp.                          &       &      &          &           & B$_i$: 0.57(5)\\
                              &       &      &          &           & B$_s$: 1.15(10)\\
\hline
\hline
\end{tabular}
\newline
$^\dagger$ PAW calculations.\\
$^\ddagger$ Unstable, converts to $s\langle100 \rangle$ dumbell during relaxation. \\
\label{tbl_bsi}
\end{table}

\subsection{Copper--boron solid solution}

As noted in the Introduction, the available observations do not allow one to conclude unambiguously 
whether the ground state\cite{note:ground} of boron in bulk Cu is interstitial \bi\ or substitutional \bs. 
We calculated the heats of solution of the both impurity types using 32 and 108 atom supercells   
and find that 
\bi\ (in octahedral site) is marginally more stable (by 0.04 eV with a 108 atom supercell, Table~\ref{tbl_bsi}). 
The difference is small hence it seems reasonable to expect that both \bi\ and \bs\ can be found
at elevated temperatures, as was indeed observed.\cite{ittermann96,stockmann01,fullgrabe01}     

However, as shown in Table~\ref{tbl_bsi}, either the insertion of a boron atom into
an interstice or replacement of a host atom at a regular lattice site both lead to significant
volume change, $\Omega_d$, negative for \bs\ and positive for \bi. Hence, 
one could hypothesise that combining \bs\ and \bi\ into a \textit{dimer} might eliminate most of the
elastic distortion of the lattice. To explore this idea, we repeated the calculation with 
\bi\ and \bs\ placed next to each other either along the $\langle111\rangle$ or $\langle100\rangle$ directions.
In addition, we also considered two boron atoms arranged symmetrically around a vacant site 
along same directions. We shall refer to the former and the latter as asymmetric and 
symmetric dumbells, respectively. Among these, the symmetric $\langle100\rangle$ dumbell, $s\langle100\rangle$, 
appears to be 
the most stable (see Table~\ref{tbl_bsi}). The heat of solution of the $s\langle100 \rangle$ dumbell is 
by a factor of two lower than those of the single impurities, indicating that the dumbells should dominate 
at low temperatures and even survive up to the eutectic temperature $T_e = 1013$\dg.

To verify the last point, we use a simple model in which boron dumbells \bd\
together with single impurities \bi\ and \bs\ are treated as three types of coexisting 
point defects forming an ideal solution. The model is described in Appendix~A, together
with the results obtained within this model in the dilute limit. 
We find in particular, that the concentration 
of dumbells exceeds those of single impurities in most conditions unless the temperature
is close to $T_e$ or the boron content is very small. Otherwise, all three impurity forms 
coexist, both in the single phase and the two phase regions of the Cu--B phase diagram,  
with the dumbells usually being the dominant kind. 

In fact, this can be anticipated already from the
difference of the solution enthalpies \textit{per entity} ($H_s$ for single impurities and 
$2H_s$ for dumbells) listed in Table~\ref{tbl_bsi} in column $\Delta H_s$. These differences serve to
estimate the relative amount of defects at terminal solubility (see Eq.~(\ref{cm}) in the Appendix).



The fact that adding B to Cu gives rise to three types of competing defects 
is due to a remarkable coincidence that B atoms are so perfectly ``incompatible'' with the Cu
lattice that the interstitial and substitutional sites are nearly degenerate in energy;         
and furthermore, boron dumbells stabilised by this misfit strain turn out to have almost 
the same heat of solution per dumbell as those of the single impurities per atom.  

The individual concentrations of the defects might of course change if temperature
effects, such as atomic vibrations and lattice expansion, are fully taken into account.  
In addition, association of several (three, four, etc.) impurities can also play a r\^ole,
at least at low temperatures. 
Nevertheless, we believe that our results provide a strong indication that in equilibrium 
copper--boron alloys a significant fraction of the impurities are found in ``bound'' states. 

We are not aware of any metallic system in which dilute impurity would aggregate. 
Interestingly, Dewing noted that activity measurements of B in \textit{molten}
Cu suggested that B should dimerise in dilute solutions.\cite{dewing90} Here we seem 
to arrive at the same conclusion although for different reasons
as elastic strain does not exist in liquids. 
Our finding might still be relevant to the processing of the experimental 
data on copper--boron melts as the \textit{solid} alloy in these studies is customarily 
assumed to be an ideal\cite{rexer70} or regular\cite{jacob00} solution of fully dissociated 
impurity atoms. 
 
Another observation that stems from the heats of solution in Table~\ref{tbl_bsi} is 
the fact that the solubility limits indicated in the experimental phase diagram,\cite{Massalski}   
0.06 at.\% at room temperature and 0.29 at.\% at $T_e$, are much too high in comparison with
the enthalpies that we obtained. Assuming that impurity atoms are in the form 
of dimers and that boron precipitates as the pure rhombohedral $\alpha-$phase, the
above solubilities would translate into the Gibbs free energy of solution 0.42 eV/atom at $T_e$
and 0.12 eV/atom at room temperature. Assuming single impurities leads to even larger disagreement.
%
%
Using the aforementioned ideal solution model 
together with the heats of solution in Table~\ref{tbl_bsi} leads to 
a three orders of magnitude discrepancy in terminal solubility at $T_e$, 
which increases to more than twenty orders of magnitude at room temperature (see Appendix A).
Temperature effects can be noticeable at $T_e$ but cannot explain 
the discrepancy at room temperature. 
   
Puzzled with this inconsistency, we compared our FP LMTO
calculations with those by the PAW method\cite{blochl1994,kresse1999} as implemented in the VASP 
code.\cite{kresse1993,kresse1996} $H_s$ obtained for \bi\ in the 32 atom cell, also shown in 
Table~\ref{tbl_bsi}, agrees with the LMTO result within 0.05 eV/atom (the difference almost 
entirely comes from the energy difference between fcc and $\alpha-$boron obtained by each method, 
see Table~\ref{tbl_balpha}).  
Hence we conclude that the heats of solution presented in Table~\ref{tbl_bsi} are
the correct LDA result.

The B--Cu phase diagram in Ref.~\onlinecite{Massalski} is taken from the critical assessment
of available experimental data by Chakrabarti and Laughlin,\cite{chakrabarti82} 
in which the data on maximal solubility of B in Cu are solely based on experimental
work by Smiryagin and Kvurt.\cite{smiryagin65} The solubility limits mentioned above
are those estimated in this latter study (0.05 and 0.01 wt.\% B translated into at.\%) and 
appear to serve more as an upper boundary rather than as exact numbers. Chakrabarti and 
Laughlin indeed comment in their assessment that ``it is likely that the actual solubility 
is even lower than that given by [\onlinecite{smiryagin65}].'' We expect it to be
\textit{significantly} lower and appeal to future experimental work to correct the terminal boron
solubility in published B--Cu phase diagrams. 

\begin{table*}
\caption{$\Sigma 5$(310)[001] Cu grain boundary with B at various segregation sites: grain boundary excess volume per
unit area, $v^{xs}$, average segregation energies per impurity atom, $\Eseg$, and the work of separation, $\Wsep$.
Letters ``l'', ``t'', and ``i'' in the second line correspond to
impurity atoms being initially placed in the ``loose'', ``tight'', and ``interstitial'' positions at the grain boundary
plane (see Fig.~\protect\ref{fig_pure}) and then relaxed.
The optimal impurity distribution between cleaved surfaces is indicated in the last line.
The quantities corresponding to the lowest energy grain boundaries at each coverage are highlighted in bold.
 }
\begin{center}
\begin{tabular}{|l|c|ccc|ccc|c|c|}
\hline
\hline
\cline{2-10}
Impurity excess, ML       & \multicolumn{1}{c|}{0} & \multicolumn{3}{c|}{0.5}
& \multicolumn{3}{c|}{1}   & \multicolumn{1}{c|}{1.5} & \multicolumn{1}{c|}{2} \\
\hline
Site                      &      &   l   &    t    &  i   &   l+t   &   l+i  &  t+i    & l+t+i & \\
\hline
GB excess volume per unit &&&&&&&&&\\
area $v^{ex}$, \AA
                          & {\bf 0.28}& 1.02 & 0.53 & {\bf 0.45}&{\bf  0.87}&    0.78&    0.66 &{\bf 0.98} & \ldots \\
~&&&&&&&&&\\
Segregation energy $\Eseg$, eV: &&&&&&&&&\\
to the $(310)$ surface    &   0  & 0.40  &  0.40   & 0.40 &  0.17   &  0.17  &  0.17   & 0.37   & 0.30 \\
to the grain boundary     &   0  &$-$0.66&  0.40   &{\bf 0.96}&{\bf 0.51}&  0.31  &  0.29   &{\bf 0.46} & \ldots \\
~&&&&&&&&&\\
Work of separation $\Wsep$, J/m$^2$
                          & {\bf 3.35} & 2.49  &  3.35   & {\bf 3.81} &{\bf 3.57}&  3.24  &  3.21   &  {\bf 3.58}    & \ldots  \\
cleavage mode             & \ldots &  \multicolumn{3}{c|}{0.5 + 0.5} & \multicolumn{3}{c|}{0.25 + 0.75} & 0.5 + 0.5  & \ldots  \\
\hline
\hline
\end{tabular}
\end{center}
\label{tbl_b}
\end{table*}

\section{Boron at a copper grain boundary}
\label{sec_gb}

\subsection{Work of separation and grain boundary excess volume}
\label{subsec_wsep}

The work of separation of a grain boundary at given impurity excess is the difference in total 
energy between equivalent pieces of material containing the boundary and free surfaces into which the 
boundary cleaves. Both the grain boundary and the surface pieces should be taken in their lowest 
energy state.   

\begin{figure}[ht]
\begin{center}
\includegraphics[width=9 cm,angle=0]{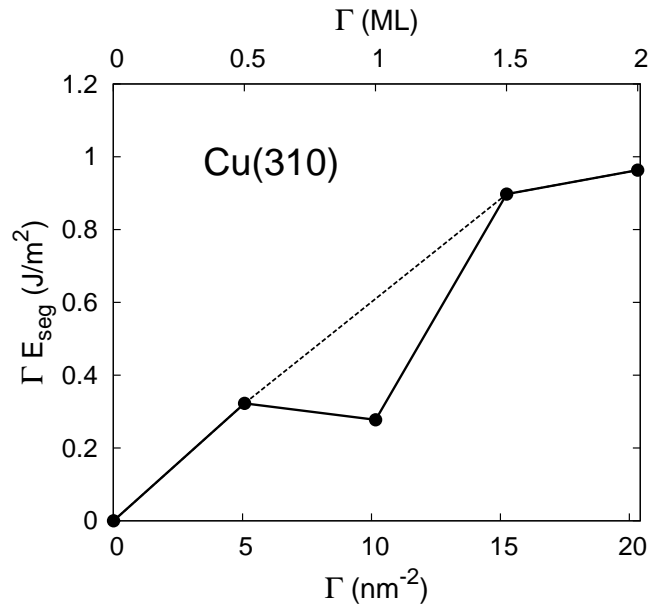}
\end{center}
\caption{Convex hull plot for B impurity at Cu(310) surface. Circles correspond to the
lowest energy surfaces for each coverage found in our study.  
1~ML surface appears to be unstable with respect to 
decomposition into those with 0.5 and 1.5~ML coverage (dashed line).
}
\label{fig_convex}
\end{figure}

\begin{figure}[hb]
\begin{center}
\includegraphics[width=9 cm,angle=0]{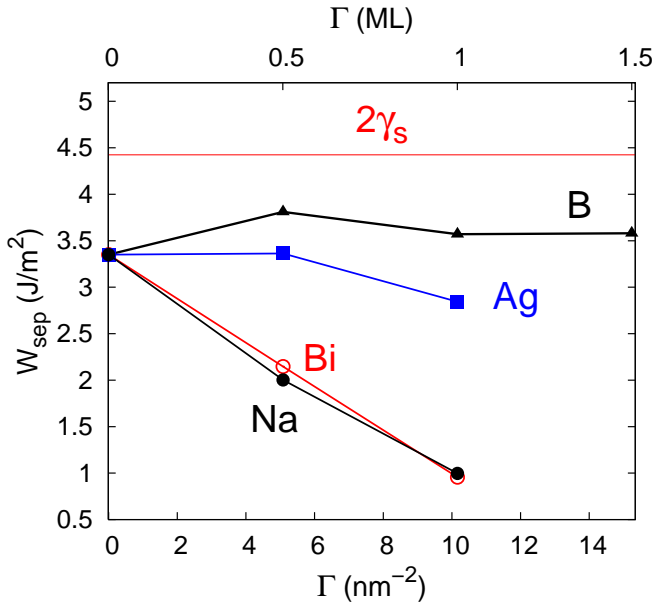}
\end{center}
\caption{The work of separation of the $\Sigma 5$(310)[001] Cu grain boundary, $\Wsep$ as
a function of impurity excess $\G$. Data for Bi, Na, and Ag are from 
Ref.~\protect\onlinecite{lozovoi06}. The horizontal line corresponds to twice 
the surface energy of Cu(310),\protect\cite{lozovoi06} $\gamma_s = 2.21$ \Jm\ 
which provides a natural upper bound for $\Wsep$.}
\label{fig_wsep}
\end{figure}

\begin{figure}[hb]
\begin{center}
\includegraphics[width=9 cm,angle=0]{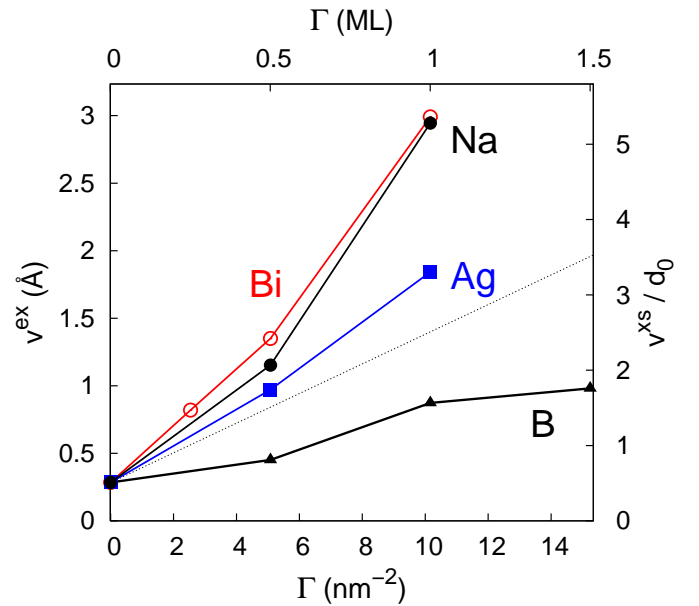}
\end{center}
\caption{Same as Fig.~\protect\ref{fig_wsep} but for the grain boundary excess
volume per unit area, $v^{xs}$. The right hand  scale gives $v^{xs}$  
in units of \{310\} interlayer spacing in bulk copper, $d_0 = 0.5576$ \AA.  
The dotted line corresponds to a hypothetical ``ideal'' impurity
identical with Cu atoms.
}
\label{fig_expan}
\end{figure}

The lowest energy grain boundary can be found by comparing the energies of relaxed grain boundaries
with impurities initially placed into various sites (substitutional or interstitial). 
This procedure does not, of course, guarantee arrival at the global minimum but is a practical alternative
to a full optimisation of grain boundary structure and includes rigid translation of the grains.   

The surfaces do not require translations, but a complication here is that one does not know in advance the
optimal distribution of the impurity atoms between two newly created surfaces. Usually the impurity 
splits equally, but not always. More generally, even if equal amount of impurity is experimentally 
detected for a cleaved {\it macroscopic} sample, there still remains a possibility that the surfaces 
would contain patches with uneven impurity coverage.  
 
The combination of surfaces which produces the lowest energy at given overall amount  
of impurity can be identified if one employs the \textit{convex hull} construction.   
In Fig.~\ref{fig_convex} we plot the total segregation energy to a surface $\Gseg^s = \G \Eseg^s$ as a function of 
impurity excess $\G$. The quantity $\Gseg^s$ shows how much the segregation decreases the energy of a piece of 
material containing a surface. Therefore, any concave region of the curve indicates that there exists a 
combination of surfaces which provide lower energy. In our case, we find that the 1~ML grain boundary 
splits into surfaces with coverage 0.5 and 1.5~ML, whereas the other boundaries cleave evenly.

The resulting works of separation are listed in Table~\ref{tbl_b}, together with segregation energies and
grain boundary excess volumes. The maximal $\Wsep$ for each coverage correspond to the lowest energy
grain boundary and are highlighted in bold. We thus find that at 0.5~ML boron prefers the interstitial 
site and at 1~ML replacing Cu at both the ``loose'' and ``tight'' sites provides the best option. 
Corresponding works of separation are compared with those obtained for Bi, Na, and Ag\cite{lozovoi06}
in Fig.~\ref{fig_wsep}. Contrary to these latter impurities, boron increases $\Wsep$ for the whole
range of coverages considered. 
The largest strengthening effect is observed at 0.5~ML at which $\Wsep$ increases by 0.5~\Jm and becomes
as large as 3.81~\Jm. This can be compared with twice the 
surface energy of pure Cu,\cite{lozovoi06}  $2\gamma^s_{310}= 4.42$~J/m$^2$ which provides 
an upper bound for $\Wsep$ above which a grain boundary would become stronger than bulk. 
 
Grain boundary segregation energies in Table~\ref{tbl_b} compare favourably with experimental estimations of 
0.4--0.5 eV,\cite{glikman74} especially for high coverages. These energies assume the $s\langle 100 \rangle$ dumbells to be
the ground state of B in bulk Cu (see Table~\ref{tbl_bsi}). Had we used interstitial or substitutional B impurities
instead, then the segregation energies would have been by 0.8 eV higher. We take this
as an independent confirmation of the fact that B impurities in bulk Cu dimerise.   

Grain boundary excess volumes $v^{xs}$ in Table~\ref{tbl_b} are calculated using the dilute bulk limit, 
Eq.~(\ref{vxs}). They are smaller than those for Bi, Na, or Ag as shown in Fig.~\ref{fig_expan}. 
The dotted line in Fig.~\ref{fig_expan} corresponds to the excess volume if Cu is notionally considered 
as an impurity (one could think of $^{65}$Cu isotope, for instance). Segregation of such ``ideal'' impurity 
leaves any grain boundary intact, hence, in accord with Eqs.~(\ref{Gamma})--(\ref{vxs}), $v^{xs}(\G)$ is a straight 
line with the slope given by $\Omega_0$.  
The fact, that the excess volumes of grain boundaries with B lie below the dotted line in Fig.~\ref{fig_expan}
indicates a denser packing of atoms at the grain boundary with boron compared to the pure boundary.  
  
\begin{figure*}[ht]
\begin{center}
\includegraphics[width=14 cm,angle=0]{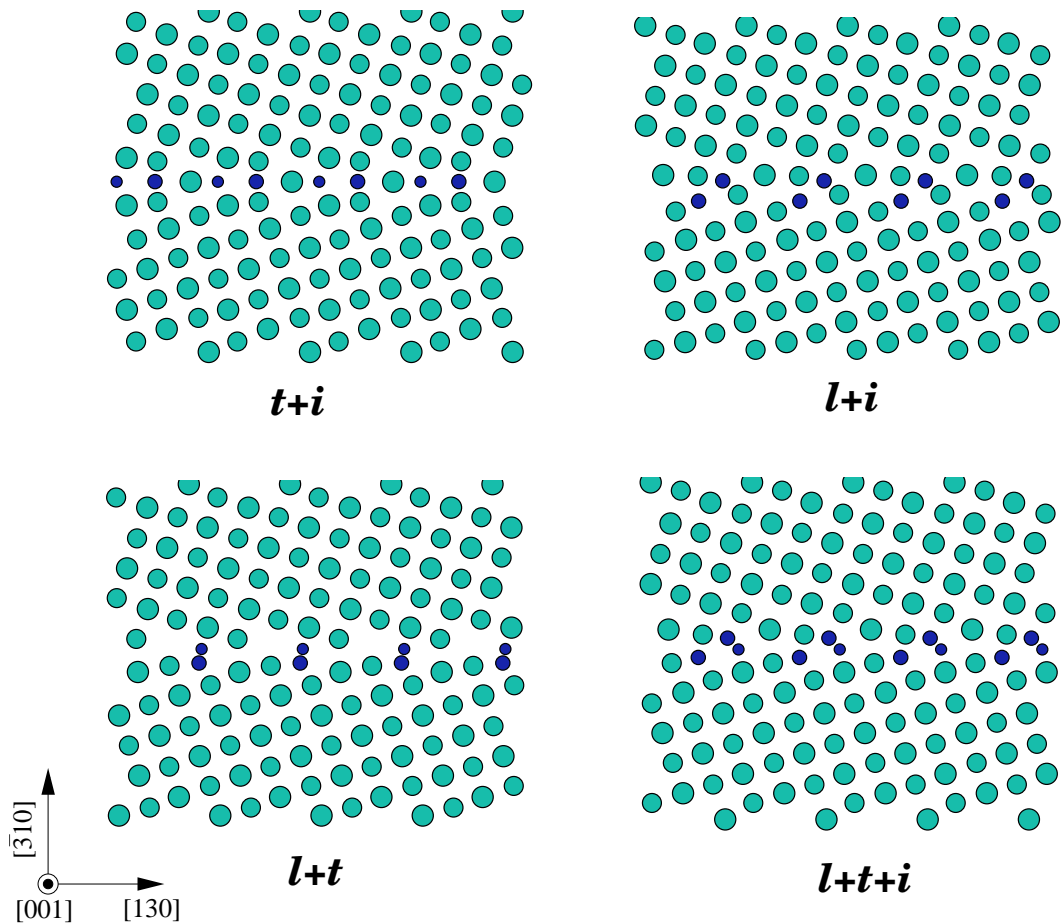}
\end{center}
\caption{Relaxed $\Sigma 5$ copper grain boundary with 1 and 1.5 ML of boron. For 1.5 ML, boron
occupies ``loose'', ``tight'', and ``interstitial'' sites at the grain boundary plane. For 1 ML, any
two of them are occupied (three combinations). Works of separation and grain boundary excess volumes   
corresponding to these boundaries are listed in Table~\ref{tbl_b}.
Larger and smaller circles correspond to host and impurity atoms lying in neighbouring (001) planes.
}
\label{fig_bgb}
\end{figure*}

\begin{figure*}[ht]
\begin{center}
\includegraphics[width=14 cm,angle=0]{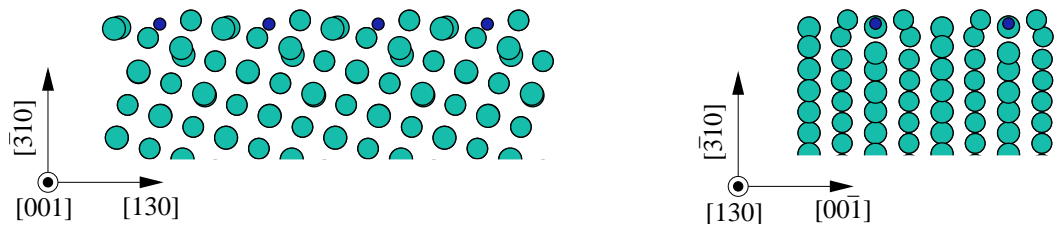}
\end{center}
\vspace{-0.5 cm}
\caption{Two side views of the relaxed Cu(310) surface with 0.5 ML of boron.
Boron atoms were initially placed instead of the top layer of Cu atoms from which
every other atom along the [001] direction was removed. During the relaxation, boron 
atoms descend below the next Cu layer causing a noticeable distortion of the latter.
}
\label{fig_bs05}
\end{figure*}

\subsection{Atomic structure of the segregated surfaces and grain boundaries}
\label{subsec_astruc}

The atomic structure of relaxed grain boundaries with 0.5~ML of boron is relatively simple.
The segregated boundaries retain the structure of the equilibrium pure boundary shown
in Fig.~\ref{fig_pure} responding to the insertion of impurity by either minor shrinking 
(``tight'' site) or expansion (``loose'' and ``interstitial'' sites). These boundaries are 
not shown here.

The structure of grain boundaries with 1 and 1.5~ML are more complex (see Fig.~\ref{fig_bgb}).  
The exception is the ``t+i'' boundary (\textit{i.e.} boron atoms segregate to the ``tight''  and 
``interstitial'' sites) which does not significantly change compared to the pure boundary.
However, the average segregation energy of boron to the ``t+i'' boundary is lower than those for 
either ``t'' or ``i'' 0.5~ML boundaries (see Table~\ref{tbl_b}).  

Among the 1~ML boundaries, the lowest energy corresponds to the more open  
``l+t'' boundary in which boron replaces copper at both ``loose'' and ``tight'' sites. 
This boundary experiences a large lateral shift of the grains. As a result of 
this shift, the boron atoms lying in adjacent (001) planes become nearest neighbours  
and form boron ``strings'' running along the [001] direction (normal to the plane of the 
drawing in Fig.~\ref{fig_bgb}). 

The ``l+i'' boundary also entails a rigid translation of the grains.
This time, however, boron atoms lie in the same plane and therefore cannot form [001] ``strings''.     
The energy of the ``l+i'' boundary is close to the energy of the ``t+i'' boundary despite 
their atomic structures being very different. As a matter of fact, the
``l+i'' boundary can be interpreted as the ``i'' boundary in which the grain boundary plane 
is shifted normal to itself by one layer and another boron atom substitutes a copper atom in the adjacent 
plane. 

The 1.5~ML ``l+t+i'' boundary combines the features of the ``t+i'' boundary (interstitial boron 
surrounded by six Cu atoms) and the ``l+t'' boundary (boron ``strings''). The average segregation 
energy for this boundary is close to that of the 1~ML ``l+t'' boundary. 
 
The following observation is worthwhile here.
Substantial fall off of segregation energy with the amount of segregant appears to be a common  
feature of boron doped intermetallic compounds. As Lej\v{c}ek and Fraczkiewicz note,\cite{lejcek03} 
this could be formally described by either introducing a strong repulsive term into the Fowler--Guggenheim
segregation isotherm, or by using the standard Langmuir--McLean isotherm but with limited number of
segregation sites. As we observe here, the former approach might be physically misleading, at least
for the Cu--B system. For instance, the 1~ML ``l+t'' boundary containing neighbouring boron 
atoms is lower in energy than the ``t+i'' boundary where boron atoms are well separated.         

The structure of free (310) surfaces with segregated boron were obtained 
by replacing copper with boron in the top layer(s) and allowing the 
surface to relax. For fractional coverages 0.5 and 1.5~ML, we tried either to substitute half of the host 
atoms in a layer with impurity (substitutional positions) or to place half a monolayer of impurity 
atoms above the top surface layer (adatom positions). The latter resulted in lower energy 
configurations.
 
During relaxation boron atoms embed into the substrate, often going beneath the top copper layer.
As a result, the top copper layer becomes strongly distorted.  Fig.~\ref{fig_bs05} shows 
the Cu(310) surface with 0.5 ML of B (and 0.5 ML of vacancies) in the first layer. 
The plane of B atoms indeed resides below the top layer of Cu atoms and
just slightly above the second layer, whereas the host atoms in the surface layer are strongly displaced 
towards the nearest boron atom.       

\begin{table*}
\caption{Work of separation $\Wsep$ at the vertices of the ``ghost impurity cycle'' (Fig.~\ref{fig_paths})
and its change $\Delta \Wsep$ due to transitions between the vertices. The latter have the meaning of the
contributions of the SS, HR, and CC mechanisms. The positive sign of $\Delta \Wsep$ corresponds to cohesion
enhancement.}
\begin{center}
\begin{tabular}{lccccccccccc}
\hline
\hline
Impurity & Site & Excess, & \multicolumn{4}{c}{$\Wsep$, J/m$^2$} & \multicolumn{1}{c}{\mbox{\hspace*{0.1cm}}} &
\multicolumn{4}{c}{$\Delta \Wsep$, J/m$^2$} \\
\cline{4-7}
\cline{9-12}
         &      &   ML    &  A  & B  &   C   &  D  &
                          &  Total  & SS  &   HR   &  CC \\
\hline
~\\
B   & i     & 0.5 &  3.35 &  3.81 &  3.38 &  3.35 & &  0.46 &  0.03 &   0     &  0.43 \\
    & l     & 0.5 &  -"-  &  2.49 &  2.20 &  2.14 & &$-$0.86&  0.06 & $-$1.21 &  0.29 \\
    & t     & 0.5 &  -"-  &  3.35 &  2.87 &  1.92 & &  0.00 &  0.95 & $-$1.43 &  0.48 \\
    & l+t   & 1.0 &  -"-  &  3.57 &  3.08 &  1.35 & &  0.22 &  1.73 & $-$2.00 &  0.49 \\
    & l+i   & 1.0 &  -"-  &  3.24 &  2.84 &  2.18 & &$-$0.11&  0.66 & $-$1.17 &  0.40 \\
    & t+i   & 1.0 &  -"-  &  3.21 &  3.15 &  1.96 & &$-$0.14&  1.19 & $-$1.39 &  0.06 \\
    & l+t+i & 1.5 &  -"-  &  3.58 &  3.33 &  1.38 & &  0.23 &  1.95 & $-$1.97 &  0.25 \\
~\\
Bi  & l     & 0.5 & -"-  & 2.15 & 1.81 & 2.20 &
                  & $-$1.20 & $-$0.39  &$-$1.15 &  0.34 \\ 
    & l+t   & 1.0 & -"-  & 0.96 & 0.13 & 1.38 &
                  & $-$2.39 & $-$1.25 & $-$1.97 &  0.83 \\ 
~\\
\hline
\hline
\end{tabular}
\end{center}
\label{tbl_w}
\end{table*}

\begin{table*}
\caption{Same as Table~\ref{tbl_w} but for the segregation energies $\Eseg$.
The latter are \textit{average} segregations energies per impurity atom (or vacancy).
The sign convention is that positive $\Eseg$ means that impurity wants to segregate.
Segregation energies for the pure boundary $\Eseg(\textrm{A})$ are set to zero.
The bulk reference used here is an interstitial impurity \bi, not a dumbell \bd\ as in Table~\ref{tbl_b}.}
\begin{center}
\begin{tabular}{lccccccccccccccccc}
\hline
\hline
Impurity & Site & Excess, &
\multicolumn{6}{c}{$\Eseg$, eV} &  \multicolumn{1}{c}{\mbox{\hspace*{0.1cm}}} &
\multicolumn{8}{c}{$\Delta \Eseg$, eV} \\
\cline{4-9}
\cline{11-18}
         &      &   ML
& \multicolumn{2}{c}{B}  & \multicolumn{2}{c}{C} &  \multicolumn{2}{c}{D} &
& \multicolumn{2}{c}{Total}  & \multicolumn{2}{c}{SS} &  \multicolumn{2}{c}{HR} & \multicolumn{2}{c}{CC} \\
\cline{4-9}
\cline{11-18}
&&&
surf & gb &
surf & gb &
surf & gb &&
surf & gb &
surf & gb &
surf & gb &
surf & gb \\
\hline
~\\
B& i    & 0.5 & 1.18 & 1.74 & $-$0.14 & $-$0.10 & 0       &   0    && 1.18 & 1.74 & $-$0.14 & $-$0.10 & 0       &   0     & 1.32 & 1.84 \\
 & l    & 0.5 & 1.18 & 0.12 & $-$0.14 & $-$1.56 & 0       &$-$1.49 && 1.18 & 0.12 & $-$0.14 & $-$0.07 & 0       & $-$1.49 & 1.32 & 1.68 \\
 & t    & 0.5 & 1.18 & 1.18 & $-$0.14 & $-$0.73 & 0       &$-$1.76 && 1.18 & 1.18 & $-$0.14 &    1.03 & 0       & $-$1.76 & 1.32 & 1.91 \\
 & l+t  & 1.0 & 1.16 & 1.29 & $-$0.34 & $-$0.50 & $-$0.02 &$-$1.25 && 1.16 & 1.29 & $-$0.32 &    0.75 & $-$0.02 & $-$1.25 & 1.50 & 1.79 \\
 & l+i  & 1.0 & 1.16 & 1.09 & $-$0.34 & $-$0.65 & $-$0.02 &$-$0.74 && 1.16 & 1.09 & $-$0.32 &    0.09 & $-$0.02 & $-$0.74 & 1.50 & 1.74 \\
 & t+i  & 1.0 & 1.16 & 1.07 & $-$0.34 & $-$0.46 & $-$0.02 &$-$0.88 && 1.16 & 1.07 & $-$0.32 &    0.42 & $-$0.02 & $-$0.88 & 1.50 & 1.53 \\
 & l+t+i& 1.5 & 1.15 & 1.25 & $-$0.41 & $-$0.42 & $-$0.03 &$-$0.84 && 1.15 & 1.25 & $-$0.38 &    0.42 & $-$0.03 & $-$0.84 & 1.56 & 1.67 \\
~\\
Bi & l     & 0.5 & 3.11 & 1.63 &  1.18 & $-$0.71 & 1.20 & $-$0.22 &
                 & 3.11 & 1.63 & $-$0.02 & $-$0.49 & 1.20 & $-$0.22 & 1.93 &  2.34 \\ 
   & l+t   & 1.0 & 3.01 & 1.54 &  1.23 & $-$0.75 & 1.22 &  0.02 &
                 & 3.01 & 1.54 &  0.01 & $-$0.77 & 1.22 &  0.02 & 1.78 &  2.29 \\ 
~\\
\hline
\hline
\end{tabular}
\end{center}
\label{tbl_es}
\end{table*}

\subsection{The reasons behind grain boundary strengthening}
\label{subsec_bmech}

We now apply the ``ghost impurity cycle'' to the grain boundaries described in the previous two subsections
in order to understand why boron segregation has positive effect on $\Wsep$.
Contributions from HR, SS, and CC mechanisms defined in Sec.~\ref{sec_cycle} in terms of the work of 
separation and segregation energies are listed in Tables~\ref{tbl_w} and \ref{tbl_es}, respectively. 
We choose to evaluate the segregation energies in Table~\ref{tbl_es} assuming interstitial boron B$_i$ to be
the bulk ground state. With this choice, we can directly compare contributions to the SS mechanism 
with those obtained in other studies, Table~\ref{tbl_b_i}. 
The segregation energies for configurations \C\ and \D\ become just energies
required to create unrelaxed vacancies at an interface taken with opposite sign 
(our convention here and in Ref.~\onlinecite{lozovoi06} is that the bulk impurity is always fully 
relaxed, whether this is a real impurity or a vacancy; relaxed ``interstitial vacancy'' is just 
perfect bulk). 
If one wants to change to boron dumbells B$_d$, then the segregation energies in Table~\ref{tbl_es} 
should be modified as follows. $\Eseg$       
for configuration \B\ decreases by 0.78~eV (the difference between the enthalpies of solution
of B$_i$ and B$_d$) and $\Eseg$ for configurations \C\ and \D\ decrease by 1.27/2~eV 
(half the vacancy formation enthalpy in pure bulk, Ref.~\onlinecite{lozovoi06}). 
Works of separation in Table~\ref{tbl_w} do not depend on the bulk reference as it cancels
out in Eq.~(\ref{Wseg}).

Intuitively, one may expect that segregation of boron to interstitial sites at 0.5~ML coverage
would reinforce the boundary because additional atoms 
lead to additional cohesion across the interface provided that the boundary is not much distorted. 
Table~\ref{tbl_w} supports this expectation as the total increase of $\Wsep$ by 0.46~\Jm\ is 
provided almost exclusively by the CC mechanism. The SS contribution is also positive but 
small whereas the HR contribution for interstitial impurity is zero by definition. 
Two other 0.5~ML configurations, ``l'' and ``t'', do not increase $\Wsep$   
because of the large negative HR contribution arising if boron replaces host atoms.

\begin{table*}
\caption{Boron impurity at an interstitial grain boundary site in different materials: 
contributions of the SS and CC mechanisms in terms of the difference of segregation energies 
$\Eseg^{gb} - \Eseg^{s}$ (in eV per impurity atom). For the SS mechanism, individual surface
and grain boundary contributions, 
$\Eseg^{s}$ and $\Eseg^{gb}$, are also shown. 
The energies in  Refs.~[\onlinecite{chen05,braithwaite05,wu96,geng99,janisch03}] 
are defined through the difference of binding rather than segregation energies. The results, however, 
can be compared directly.
Note that the SS and CC mechanisms in Refs.~[\onlinecite{geng99}] and [\onlinecite{janisch03}]  
are referred to as the \textit{mechanical} and \textit{chemical} 
contributions, respectively. 
}
\begin{center}
\begin{tabular}{lccccccccc}
\hline
\hline
Host & Grain boundary & Method & Total &  \multicolumn{3}{c}{SS} & CC & Cohesion & Ref. \\
\cline{4-4}
\cline{5-7}
\cline{8-8}
     &                &        &gb -- surf& surf & gb & gb -- surf&gb -- surf & enhancer? &\\ 
\hline
~\\
Fe & $\Sigma 5$(010)[001]      & DMol    (LDA) & 1.96 & \ldots  &\ldots  &\ldots &\ldots & yes & [\onlinecite{chen05}]  \\
Fe & $\Sigma 5$(210)[001]      & USPP    (LDA) & 0.49 & \ldots  &\ldots  &\ldots &\ldots & yes & [\onlinecite{braithwaite05}] \\
Fe & $\Sigma 3$(111)[1\mone 0] & FLAPW   (LDA) & 1.07 & \ldots  &\ldots  &\ldots &\ldots & yes & [\onlinecite{wu96}]  \\
Ni & $\Sigma 5$(210)[001]      & FLAPW   (GGA) & 0.49 & $-$0.27 &$-$0.16 & 0.11  & 0.38  & yes & [\onlinecite{geng99}]\\
Mo & $\Sigma 5$(310)[001]      & MBPP    (LDA) & 2.09 & $-$0.95 &$-$0.23 & 0.71  & 1.37  & yes & [\onlinecite{janisch03}]\\
Cu & $\Sigma 5$(310)[001]      & FP LMTO (LDA) & 0.56 & $-$0.14 &$-$0.10 & 0.04  & 0.52  & yes & present\\
   &                           &               &      &         &        &       &       &     & study\\
\hline
\hline
\end{tabular}
\end{center}
\label{tbl_b_i}
\end{table*}

A similar effect of interstitial boron is found in \ab\ studies of grain boundaries
in Fe, Ni, and Mo (see Table~\ref{tbl_b_i}). Boron improves the cohesion at all grain 
boundaries, and this is  mostly due to the CC mechanism. The SS contribution enhances 
grain boundary strength even more despite the surface and grain boundary terms being themselves
negative, \textit{i.e.} boron distorts free surfaces more than grain boundaries.

As boron segregation proceeds beyond 0.5~ML, the boundary is still strengthened although
the magnitude of the effect is diminished. Higher 
coverage configurations necessarily include the removal of host atoms since all interstitial
sites are already filled at 0.5~ML. Hence, the mechanism of strengthening is different.

To understand this mechanism, it is instructive to compare boron results with those for 
bismuth,\cite{lozovoi06} reproduced in the Tables for convenience.
A striking difference between B and Bi (or other oversized impurities studied in 
Ref.~\onlinecite{lozovoi06})       
is the \textit{sign} of the SS contribution. The SS mechanism always increases $\Wsep$ for 
boron (Table~\ref{tbl_w}) and decreases $\Wsep$ for Bi, Na, and even Ag without any exception. 
Compare, for example, boron and bismuth at the ``loose'' site, 0.5 ML coverage. In both cases 
the boundary is weakened, more by Bi, less by B. The negative HR contributions are similar, 
the CC mechanism acts so as to strengthen the boundary
and is even more efficient for Bi than for B. Thus, it is the SS mechanism which 
makes the difference, being positive for B but negative for Bi. The ``loose'' site
is not of course the best choice for boron, but even here it is much less harmful than Bi. 
The comparison of B and Bi for the 1 ML ``l+t'' case is even more telling. The HR 
mechanism again,
has large detrimental effect for both, the CC mechanism acts in the opposite
direction and is nearly twice as large for Bi than for B. Finally, the negative SS 
contribution is large enough to make the boundary brittle in the Bi case,\cite{lozovoi06}    
but the large and positive SS contribution for B results in the boundary being 
strengthened. In other words, \textit{the difference in the distortion pattern of grain
boundaries and surfaces} is itself sufficient to either strengthen the boundary 
or to make it brittle. 

What are the reasons for the SS contribution being positive for boron? 
Let us analyse surface and grain boundary contributions for configurations 
``l'' and ``l+t'' for B and Bi in Table~\ref{tbl_es}.
The surface contribution to the SS mechanism for Bi is negligible, therefore the
negative (embrittling) effect comes from the grain boundary distortion.
For boron, this differs in two ways. Firstly, there is always a negative surface
term, and secondly, the grain boundary term can be large and positive. Even if the 
latter is negative (as in cases ``l'' and ``i''), it is still smaller than the surface
which renders their difference positive. 

It is easy to see why the surface contribution to SS is negative for boron. 
It indicates a sizable distortion of the surface region and arises for impurities that 
can embed themselves into surface layers (Sec.~\ref{subsec_astruc}). This would be 
possible for small impurities, especially if they prefer interstitial positions in 
the bulk. 

It is less obvious why the grain boundary contribution to SS tends to be positive. 
SS contribution in the ``ghost impurity cycle'' is the energy change when a pure grain
boundary with preinserted unrelaxed vacancies \D\ is further deformed as prescribed by 
``impurity ghosts'' to arrive at configuration \C. Atoms  
in configuration \D\ would want to relax towards the vacancies, whereas deformation 
corresponding to large impurity atoms forces them to move further away. The total 
energy increases and the SS contribution is negative (embrittling). For small substitutional 
impurities this is reversed---during the \D$\to$\C\ transition atoms move 
\textit{towards} the vacancies. Hence, the energy decreases and 
SS is positive (cohesion enhancing). If boron occupies \textit{both} interstitial and substitutional sites,
atomic displacements are more complex. However, the fact that the grain boundary
excess volume is always smaller than that of the ``ideal'' Cu-like
impurity (Fig.~\ref{fig_expan}) indicates that on average the grain boundary shrinks 
rather than expands. 

The above reasoning relies only on the property of boron atoms being ``smaller''
than host atoms and therefore seems applicable to other undersize impurities,
at least for the light metalloid impurities. 
Boron was found to reinforce grain boundaries in all materials studied (Table~\ref{tbl_b_i}).
Carbon segregation increases\cite{wu96c,janisch03} or slightly
decreases  $\Wsep$,\cite{braithwaite05} 
whereas H, N, and O weaken grain boundaries.\cite{geng99,zhong00,janisch03} 
In the latter case, the embrittling propensity is due to the CC contribution, which becomes 
large and \textit{negative}.
Janisch and Els\"asser\cite{janisch03} 
suggest that this should be the case for light species whose outer electronic 
shell (1$s$ for H, 2$p$ for N and O) falls near or below the bottom of the valence band of 
the host metal.
  
Negative CC indicates that the insertion of an impurity atom into a prepared hole 
(configuration \C) weakens atomic bonds across the interface. This could be the
case if the impurity affects the bonds between the neighbouring host atoms by means 
of withdrawing electronic charge from them---which is known as the \textit{electronic} 
mechanism of embrittlement. An alternative explanation, advocated in Ref.~\onlinecite{janisch03},
is Cottrell's ``unified theory'' which refers to the position of the impurity levels
relative to the Fermi energy of the host metal.\cite{cottrell90} According to this theory, 
interstitial impurities whose valence electrons lie close to the Fermi level, would form predominantly 
covalent bonds with host atoms and hence prefer grain boundaries over surfaces due to a higher 
coordination in the former (the Cottrell $\sqrt{z}$ factor). That means positive CC and 
cohesion enhancement. On the other hand, impurities with valence states lying high above 
or deeply below the Fermi level would form polar bonds with the host atoms and turn into screened 
ions, for which the surface environment is more favourable. This results in a negative 
CC contribution which weakens the boundary.   

In our recent calculations of Cu grain boundary with inert gas atom impurities 
(He and Kr) we also observed a large negative CC contribution leading to catastrophic 
embrittlement.\cite{inert} As no charge transfer to or from inert gas atom is expected, 
the embrittling effect in this case must be related to the Pauli exclusion principle. 
One may, therefore, hypothesise that a similar mechanism can act for impurities 
with nearly completed $p$-shell, such as fluorine, oxygen and, to lesser extent, nitrogen, 
as filling the impurity shells with metal electrons would
effectively render the dopant atoms inert gas like.
The question as to how significant this ``inert gas atom'' mechanism is in comparison with others 
requires a separate investigation and is outside the scope of the present 
study.      


\section{Conclusions}
\label{sec_concl}

The effect of boron impurities at the
$\Sigma 5$(310)[001] grain boundary, (310) surface and in the bulk of Cu is 
investigated on the 
basis of first principles calculations using the full potential LMTO method.
 
1. We find that B strengthens the boundary in the whole range of coverages 
studied (up to 1.5~ML) with the maximal effect achieved at 0.5~ML.
Combined with the observed ability of B to remove harmful impurities such as 
Sb from the copper boundary,\cite{glikman74} this makes boron a particularly attractive
alloying addition.     

2. The reasons behind grain boundary strengthening at 0.5~ML and higher
coverages are different. 0.5~ML corresponds to all interstitial positions 
at the boundary being filled by boron atoms providing therefore additional
cohesion between the grains while not distorting the boundary much (the CC mechanism). 

3. At 1 and 1.5~ML boron begins to substitute host atoms at the boundary leading
to significant distortions and lateral translations of the grains. The SS contribution,
however, remains positive and acts so as to increase $\Wsep$. We demonstrate that
the difference in the sign of the SS contribution proves to be 
solely responsible for the opposite effect of B compared to embrittling species
such as Bi. 

 
4. Distortion of a free surface by segregated boron atoms further increases
$\Wsep$, but it is not a decisive factor. 

5. Introducing boron into bulk Cu leads to a peculiar situation in which substitutional
and interstitial impurities are rather close in energy. Combined together, they form
a strongly bound dimer held by elastic forces of the host lattice. 
Remarkably, the heat of solution of the lowest energy $s\langle 100 \rangle$ dumbell (per dimer)  
is also close to the heat of solution of boron single impurities (per atom). 
Thus a sizable proportion of boron atoms should be found in a bound state
in most experimental conditions, even at high temperature.

6. A large discrepancy between calculated heats of solution and experimental estimations
for terminal solubility of B in Cu is discovered. We are inclined to think that the 
solubility limits suggested in Ref.~\onlinecite{smiryagin65} and then translated into 
existing  B--Cu phase diagram, are overestimated by a few orders
of magnitude and hope that our findings inspire experimental work on 
the updated version of the  phase diagram.

\section*{Acknowledgements}

The work has benefited from salutory discussions with M.~W. Finnis, to
whom we are also grateful for a number of valuable suggestions to the manuscript.
Useful comments from M.~van Schilfgaarde and P.~Ballone are much appreciated. 
We thank T.~P.~C. Klaver for an independent PAW calculation of pure boron and
boron impurity in bulk copper. 

\appendix
\section{Equilibrium concentration of coexisting impurity types in an ideal solution}

We describe a thermodynamic approach which we use in the paper to estimate the 
equilibrium concentration of boron single impurities (\bi\ and \bs) and dimers (\bd) in bulk Cu. 
The approach follows the canonical treatment of ideal solid solutions proposed in Ref.~\onlinecite{hagen1998} 
and is applicable to any binary system in which the heats of solution of the impurity species 
occupying an interstitial position, a substitutional position, or forming a dumbell, 
are comparable.


The reader may be surprised by the fact that the model outlined below ignores thermal vacancies  
altogether. Indeed, at a first glance this looks inconsistent given that the heats of solution of boron 
in Cu listed in Table~\ref{tbl_bsi} are comparable with the vacancy formation enthalpy 
$H_f^v = $1.27 eV.\cite{lozovoi06} We omit vacancies deliberately as the equilibrium impurity concentrations 
\textit{do not depend} on the vacancy formation enthalpy, hence they do not change even if no vacancies 
at all are allowed. Indeed, if a crystal contains thermal vacancies with concentration $c_v$, an additional
term will appear in Eqs.~(\ref{site}), (\ref{E}), and (\ref{S}), and, consequently, one more relation 
will be added to the system of equations (\ref{sys}). However, $c_v$ can be eliminated from (\ref{sys}) 
explicitly leading to exactly the same set of equations (\ref{si})--(\ref{c}) as below.  
The equilibrium impurity concentrations, obtained as the solution of Eqs.~(\ref{si})--(\ref{c}),  
will therefore not depend on the vacancy concentration either. (Note in passing that 
the reverse is not true, \textit{i.e.} the equilibrium concentration of
vacancies does depend on impurity concentrations.)   

\subsection{Variables and definitions}
\label{sec_def}

Consider a large piece of
crystal $A_{1-x}B_x$ containing $N$ lattice sites with $n_a$ host atoms $A$ and $n_b$ impurity atoms $B$. 
The latter in turn, include $n_s$ substitutional impurities, $n_i$ interstitial impurities, and
$n_d$ dumbells:
$$
n_s + n_i + 2n_d = n_b \, .
$$
If the crystal is sufficiently large so that any surface effects can be neglected, the resulting 
equilibrium concentrations should depend on $n_a$ and $n_b$ only through the composition 
of the solid solution 
\beg \label{x} 
x = \frac{n_b}{n_a+n_b} \, . 
\e 

Concentrations of host atoms ($c_a$) and impurity of any type ($c_s, c_i$,  and $c_d$),
defined ``per lattice site'' here, must satisfy the following two constraints:  
\begin{eqnarray}
N c_a &=& n_a \label{na} \\
N \left(c_s + c_i + 2c_d\right) &=& n_b \, , \label{nb} 
\end{eqnarray}
indicating that the total number of atoms of each species is conserved. In addition, 
we have the ``site balance'' condition as every lattice position should be occupied
by either a host atom, an impurity atom, or an impurity dumbell: 
\beg
\label{site}
c_a + c_s + c_d = 1 \, .
\e
Overall, there are five variables ($N, c_a, c_s, c_i$,  and $c_d$) and three constraints
(\ref{na})--(\ref{site}), hence the system has two degrees of freedom. 
  

If defects do not interact, the total energy of the system is linear in defect 
concentrations: 
\beg
\label{E}
E = N \left( c_a \varepsilon_a + c_s \varepsilon_s + c_i \varepsilon_i + 2 c_d \varepsilon_d \right) \, ,
\e
where $\varepsilon_a$ is the energy (per atom) of the pure crystal, whereas  $\varepsilon_s, \varepsilon_i,$ and $\varepsilon_d$ 
are the energies per impurity atom defined by Eq.~(\ref{E}). In practice these are usually found
from total energy calculation of supercells containing a single defect of each type. 
Enthalpies of solution $H_s$, such as those listed in Table~\ref{tbl_bsi}, are related to the $\varepsilon$'s
in a simple way:
$$ 
H_s^i = \varepsilon_i - \varepsilon_b; \quad 
H_s^s = \varepsilon_s - \varepsilon_b; \quad 
H_s^d = \varepsilon_d - \varepsilon_b \, ,
$$ 
where $\varepsilon_b$ denotes the energy per atom of species $B$ in its pure state. 

 
At zero pressure, the Gibbs free energy of the crystal is 
\beg
\label{G}
G = E - TS \, ,
\e
where $T$ is the temperature and $S$ is the (configurational) entropy given by 
\begin{eqnarray}
\label{S}
S &=& - k N \Big[ 
c_a \log c_a + c_s \log c_s + c_d \log c_d - c_d \log \eta \nonumber \\
&+& c_i \log c_i
+ \left(\alpha - c_i \right) \log \left( \alpha - c_i\right) - \alpha \log \alpha \Big] \, ,
\end{eqnarray}
where $k$ is the Boltzmann constant, $\eta$ is the number of equivalent orientations of 
the dumbell, and $\alpha$ is the number of interstitial sites per one lattice site. For 
the octahedral interstices in the fcc lattice $\alpha=1$, whereas $\eta = 3$ for 
the $\langle 100 \rangle$ dumbell in cubic crystals. Eqs.~(\ref{E}) and (\ref{S}) do 
not take into account atomic vibrations, but these can be easily included, for example, 
at the level of quasi harmonic approximation.\cite{lozovoi03n}  

\subsection{Equilibrium concentrations}
\label{sec_1phase}

The equilibrium impurity concentrations are those that minimise the Gibbs free energy (\ref{G})
subject to constraints (\ref{na})--(\ref{site}). The minimisation leads to the following 
system of five equations 
\begin{eqnarray}
G - \mu_a n_a - \mu_b n_b &=& 0 \nonumber \\
\varepsilon_a + kT (1 + \log c_a) - \mu_a - \lambda/N &=& 0 \nonumber \\
\varepsilon_s + kT (1 + \log c_s) - \mu_b - \lambda/N &=& 0 \label{sys} \\
2\varepsilon_d + kT (1 + \log c_d/\eta ) - 2 \mu_b - \lambda/N &=& 0 \nonumber \\
\varepsilon_i + kT \log \left[ c_i/(\alpha - c_i) \right] - \mu_b &=& 0 \nonumber  \, ,
\end{eqnarray}
where $\mu_a$ and $\mu_b$ are Lagrange multipliers associated with Eqs.~(\ref{na}) and (\ref{nb})
and therefore have the meaning of the chemical potentials of species $A$ and $B$, respectively.  
$\lambda$ is the Lagrange coefficient corresponding to Eq.~(\ref{site}). 
Eqs.~(\ref{sys}) together with constraints (\ref{na})--(\ref{site}) are sufficient to 
determine all the unknowns.

Elimination of $\mu_a$, $\mu_b$, and $\lambda$ leaves the following two independent relations
containing only the impurity concentrations: 
\begin{eqnarray}
\frac{c_i/\alpha}{(1-c_i/\alpha)^{\alpha+1}c_s} &=& \exp \left[ - \, \frac{(\varepsilon_i - \varepsilon_s)}{kT} \right]  \label{si}\\
\frac{(c_i/\alpha)c_s}{(1-c_i/\alpha)c_d/\eta} &=& \exp \left[ - \, \frac{(\varepsilon_i + \varepsilon_s - 2\varepsilon_d)}{kT} \right] \, ,  
\label{dsi}
\end{eqnarray}
in which the reader might immediately recognise ``quasi chemical'' relations describing defect reactions. 
Eq.~(\ref{si}), in particular, corresponds
to the conversion of a substitutional impurity into an interstitial one, whereas Eq.~(\ref{dsi}) describes
the dissociation of a dumbell into an interstitial and a substitutional impurity. 

Eqs.~(\ref{si})--(\ref{dsi}) together with the relation 
\beg
\label{c}
c_s + (1-x) c_i + (2-x) c_d = x
\e
can be used to find all three concentrations $c_s, c_i$, and $c_d$. 
[Eq.~(\ref{c}) readily follows  
from constraints (\ref{na})--(\ref{site}) and the definition of $x$ (\ref{x})].   

The above consideration applies to an ideal solid solution of arbitrary composition $x$.  
If the alloy is dilute, $x \ll 1$, finding the solution of Eqs.~(\ref{si})--(\ref{c}) 
simplifies and reduces to solving the quadratic
\beg
\label{quadr}
\left[2 \eta \, e_{di}^2 - e_{si} - \alpha(\alpha+2) \right] z^2 + \left(e_{si} + \alpha + 2x \right) z - x = 0 \, ,
\e
where $e_{di}$ and $e_{si}$ denote the following exponentials  
\begin{eqnarray*}
e_{di} &=& \exp \left( - \frac{\varepsilon_d - \varepsilon_i}{kT}\right)          
    = \exp \left( - \frac{H_s^d - H_s^i}{kT}\right)  \\
e_{si} &=& \exp \left( - \frac{\varepsilon_s - \varepsilon_i}{kT}\right)
    = \exp \left( - \frac{H_s^s - H_s^i}{kT}\right) \, .
\end{eqnarray*}
Equilibrium concentrations are then obtained using the positive root of (\ref{quadr}), $z_+$, as:  
\beg
\label{ceq}
c_d = \eta \, e_{di}^2 \, z_+^2 \, ; \qquad 
c_i = \alpha \, z_+ \, ; \qquad
c_s = e_{sd} \, z_+ 
\e

\subsection{Solubility limit}
\label{sec_2phase}

As pointed out in [\onlinecite{hagen1998}], the advantage of the above approach is that it produces
not only the equilibrium concentrations of defects but also the chemical potentials of the species.  
The chemical potential of the solute, in particular, can be restored from the equilibrium concentrations    
$c_d, c_i$, and $c_s$ using any of the following equations:
\begin{eqnarray}
\mu_b 
&=& \varepsilon_s + kT \left[ \log c_s + \alpha \log (1 - c_i/\alpha)\right] \nonumber\\
&=& \varepsilon_i + kT \log \left[ c_i/(\alpha - c_i) \right] \label{mub}  \\
&=& \varepsilon_d + \frac{1}{2} kT \left[ \log c_d/\eta + \alpha \log (1 - c_i/\alpha)\right] \, . \nonumber 
\end{eqnarray}
These three relations simply express the fact that impurity atoms participating in any of the
three types of defects considered here, namely 
interstitial impurity, substitutional impurity, and impurity dumbells, are 
in equilibrium with each other, hence their chemical potentials must be equal. 

Once the chemical potentials are known, it is straightforward to find the limiting solubilities by 
considering the equilibrium between the $A$-rich and the $B$-rich phases with terminal compositions. 
The terminal compositions are those that make $\mu_a$ and $\mu_b$ in the both phases equal. 
 

For our purposes, however, it is sufficient to assume that the $B$-rich phase is a pure $B$ crystal, 
such as the rhombohedral $\alpha$-boron, with $\mu_b = e_b$.  
Using again the dilute limit, from (\ref{mub}) we obtain the maximal defect concentrations in $A_{1-x}B_x$ as 
\begin{eqnarray}
c_s^m &=& \exp \left( - \, \frac{H_s^s}{kT} \right)      \nonumber  \\
c_i^m &=& \alpha \exp \left( - \, \frac{H_s^i}{kT} \right) \label{cm}     \\
c_d^m &=& \eta \exp \left( - \, \frac{2 H_s^d}{kT} \right)  \, .  \nonumber
\end{eqnarray}
These, according to (\ref{c}), define the limiting solubility  $x^m$ as
\beg
\label{xlim}
x^{m} = c_s^m + c_i^m + 2 c_d^m \, .
\e 

\begin{figure}[ht]
\begin{center}
\includegraphics[width=9 cm,angle=0]{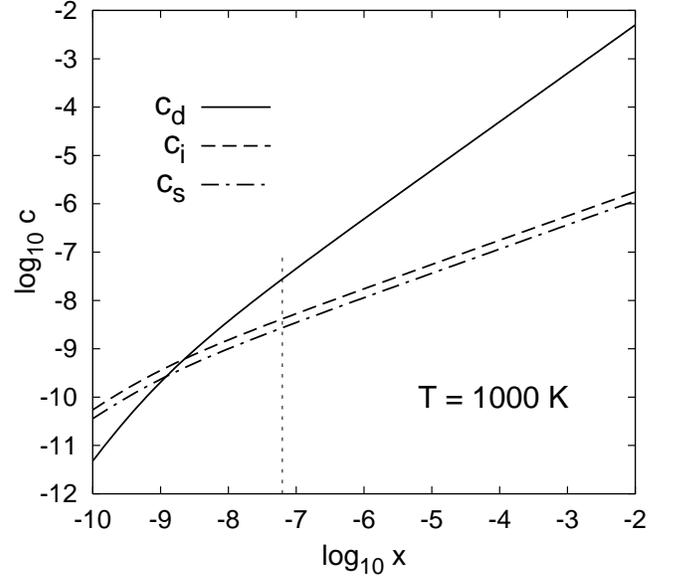}
\end{center}
\caption{
Equilibrium concentration of \bd, \bi\ and \bs\ in Cu$_{1-x}$B$_x$ at T = 1000~K 
given by Eqs.~(\ref{quadr})-(\ref{ceq}) as a function of boron content $x$. 
The vertical dotted line
corresponds to the solubility limit $x^{m}$ estimated according 
to Eq.~(\ref{xlim}).}
\label{fig_cx}
\end{figure}

\begin{figure}[ht]
\begin{center}
\includegraphics[width=9 cm,angle=0]{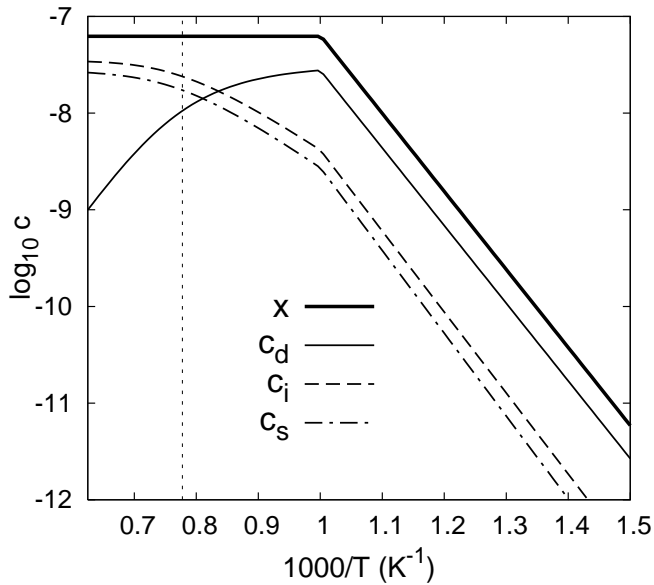}
\end{center}
\caption{
Equilibrium concentration of \bd, \bi\ and \bs\ and the total amount of boron, $x= c_i + c_s + 2c_d$, 
as a function of $T^{-1}$ (the Arrhenius plot). The alloy composition is taken as the terminal 
solution at $T = 1000$~K (vertical dotted line in Fig.~\ref{fig_cx}).
The kink on the curves corresponds to the precipitation of the 
second phase (assumed to be pure $\alpha-$B). The vertical dotted
line indicates the eutectic temperature $T_e = 1013$\dg.}
\label{fig_ct2}
\end{figure}

\subsection{Results for boron in copper}
\label{sec_apres}
 
Fig.~\ref{fig_cx} shows the equilibrium concentrations of \bi, \bs, and \bd\ at $T= 1000$K  
as a function of the boron content $x$. These were obtained using Eqs.~(\ref{quadr}) and (\ref{ceq})
with $\alpha=1, \eta=3$, and the enthalpies of solution $H_s$ listed in Table~\ref{tbl_bsi}
(those calculated in a 108 atom supercell). The limiting solubility $x^m$ given by 
Eqs.~(\ref{cm})--(\ref{xlim}) is shown with a vertical dotted line. 
 
According to Fig.~\ref{fig_cx}, the dumbells strongly prevail at and above $x^m$. (Supersaturated 
solutions may arise if the excess boron precipitates as some metastable phase, higher
in energy than the rhombohedral $\alpha$-B. Such a scenario, however, is not supported by 
experimental observations\cite{rexer70}). With $x$ decreasing, the loss of configurational
entropy should eventually overweigh the energetic advantage of forming dumbells, giving 
rise to the crossover between concentration of dumbells and single impurities. This is indeed
observed in Fig.~\ref{fig_cx},  although the crossover concentrations seem too small to  
be experimentally detectable.

Fig.~\ref{fig_ct2} shows the temperature dependence of the impurity concentrations at fixed $x$
in the form of the Arrhenius plot $\log c = f(1/T)$. Concentration $x$ for this plot 
is chosen as the limiting solubility in Fig.~\ref{fig_cx}. As a result, the curves in Fig.~\ref{fig_ct2} 
have a kink at $T=1000$~K which corresponds to the precipitation of the second phase. The curves 
to the left of the kink are the equilibrium defect concentrations in a single phase crystal Cu$_{1-x}$B$_x$
given by (\ref{quadr})--(\ref{ceq}), whereas concentrations to the right of the kink correspond 
to a two--phase equilibrium between Cu$_{1-x}$B$_x$ and pure $\alpha$-boron. The latter are given by
Eq.~(\ref{cm}) and are just straight lines in the Arrhenius coordinates.      

We again observe the dominance of boron dumbells in the whole range of temperatures, except  
the narrow region close to the eutectic temperature $T_e = 1013$\dg\ 
(vertical dotted line in Fig.~\ref{fig_ct2}). The crossover point appears here for the same
reason as in Figs.~\ref{fig_cx} and rapidly moves to higher temperatures with $x$ increasing.
If, for example, for this plot we use $x$ corresponding to the limiting solubility
at $T_e$, then the crossover would move above both the eutectic temperature  and melting temperature of pure
copper $T_m = 1085$\dg, so the dumbells would dominate everywhere.

To summarise, we observe that boron dumbells \bd\ are the essential component of Cu$_{1-x}$B$_x$ 
solid solutions both in the single phase and two phase regions of the Cu--B phase diagram. 
As a matter of fact, the concentration of boron dumbells \bd\ exceeds those of single 
impurities \bi\ and \bs\ in most conditions. High temperature and small boron content tend
to make these concentrations comparable at best. The dumbells can be suppressed only in very diluted
samples where the impurity concentrations are likely to fall below the detection limit anyway. 
 
The terminal solubility of B in Cu appears to be lower than that indicated in published phase diagrams.\cite{Massalski} 
It is of the order of a few ppm at $T_e$ (cf. 0.29 at.\% in [\onlinecite{Massalski}]), and is as 
low as $10^{-29}$ at room temperature (cf. 0.06 at.\%  in [\onlinecite{Massalski}]). The fact that
the second phase precipitates as a solid solution of Cu in B\cite{rexer70,Massalski} 
rather than as the pure boron (assumed here) should lead to even lower limiting solubilities. 

%
%

\end{document}